\documentclass{amsproc}
\usepackage{graphicx}

\newtheorem{theorem}{Theorem}[section]

\theoremstyle{definition}

\newtheorem{conjecture}{Conjecture}[section]
\theoremstyle{remark}

\numberwithin{equation}{section}

\newcommand{\CA}{{\mathcal A}}

\newcommand{\CG}{{\mathcal G}}
\newcommand{\CH}{{\mathcal H}}

\newcommand{\CK}{{\mathcal K}}
\newcommand{\CL}{{\mathcal L}}

\newcommand{\CN}{{\mathcal N}}
\newcommand{\CO}{{\mathcal O}}

\def\IZ{{\mathbb Z}}
\def\IR{{\mathbb R}}

\def\IP{{\mathbb P}}

\def\IS{{\mathbb S}}

\def\IQ{{\mathbb Q}}

\newcommand{\tr}{{\rm Tr}}
\newcommand{\re}{{\rm e}}
\newcommand{\ri}{{\rm i}}
\newcommand{\rd}{{\rm d}}

\newcommand{\pic}[2]{\raisebox{-.3\height}{\includegraphics[scale=#2]{#1}}}
\def\unknot{\pic{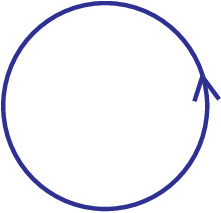} {.100}}
\def\ununknot{\pic{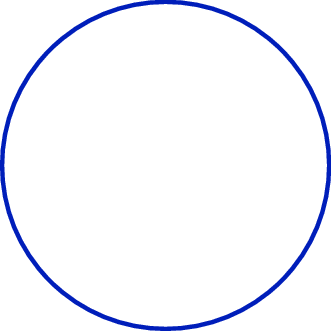} {.080}}

\newcommand{\be}{\begin{equation}}
\newcommand{\ee}{\end{equation}}
\newcommand{\ba}{\begin{aligned}}
\newcommand{\ea}{\end{aligned}}
\newcommand{\ben}{\begin{eqnarray}\displaystyle}
\newcommand{\een}{\end{eqnarray}}

\newcommand{\sectiono}[1]{\section{#1}\setcounter{equation}{0}}

\newdimen\tableauside\tableauside=1.0ex
\newdimen\tableaurule\tableaurule=0.4pt
\newdimen\tableaustep
\def\phantomhrule#1{\hbox{\vbox to0pt{\hrule height\tableaurule width#1\vss}}}
\def\phantomvrule#1{\vbox{\hbox to0pt{\vrule width\tableaurule height#1\hss}}}
\def\sqr{\vbox{%
  \phantomhrule\tableaustep
  \hbox{\phantomvrule\tableaustep\kern\tableaustep\phantomvrule\tableaustep}%
  \hbox{\vbox{\phantomhrule\tableauside}\kern-\tableaurule}}}
\def\squares#1{\hbox{\count0=#1\noindent\loop\sqr
  \advance\count0 by-1 \ifnum\count0>0\repeat}}
\def\tableau#1{\vcenter{\offinterlineskip
  \tableaustep=\tableauside\advance\tableaustep by-\tableaurule
  \kern\normallineskip\hbox
    {\kern\normallineskip\vbox
      {\gettableau#1 0 }%
     \kern\normallineskip\kern\tableaurule}%
  \kern\normallineskip\kern\tableaurule}}
\def\gettableau#1{\ifnum#1=0\let\next=\null\else
\squares{#1}\let\next=\gettableau\fi\next}

\newcommand{\figref}[1]{Fig.~\protect\ref{#1}}

\begin{document}

\title{Chern--Simons theory, the $1/N$ expansion, and string theory}

\author{Marcos Mari\~no}
\address{D\'epartement de Physique Th\'eorique et Section de Math\'ematiques, Universit\'e de Gen\`eve, CH 1211 Switzerland}

\email{marcos.marino@unige.ch}
\thanks{The first author was supported in part by FNS.}




\thanks{I am supported in part by the Fonds National Suisse. I would like to thank the organizers of the conference {\it Chern--Simons geuge theory: twenty years after} for a 
wonderful meeting.}
\begin{abstract}
Chern--Simons theory in the $1/N$ expansion has been conjectured to be equivalent to a topological string theory. This conjecture predicts 
a remarkable relationship between knot invariants and Gromov--Witten theory. We review some basic aspects of this relationship, as well as 
the tests of this conjecture performed over the last ten years. Particular attention is given to indirect tests based on integrality conjectures, both for the 
HOMFLY and for the Kauffman invariants of links. 
\end{abstract}

\maketitle


\section{Introduction}
Twenty years ago, Edward Witten provided a Quantum Field Theory (QFT) formulation of knot and three-manifold invariants \cite{witten}. The QFT is based on 
the Chern--Simons action
\be
S={1\over g_s} \int_M \tr \left(A \wedge \rd A + {2\over 3} A\wedge A \wedge A\right).
\ee
where $A$ is a connection for a gauge group $G$, and $M$ is a three-manifold. 
The partition function of this theory, 
\be
Z(M)=\int {\mathcal D} A \, \re^{\ri S(A)} 
\ee
is formally a topological invariant of $M$. This invariant can be defined rigorously and it is sometimes called the WRT (Witten--Reshetikhin--Turaev) invariant of 
$M$. Invariants of knots are obtained in the following way. Let $R$ be a representation of $G$, and let $\CK$ be a framed,  oriented 
 knot inside $M$. The holonomy of $A$ around $\CK$ will be denoted by $U_\CK$. Then, the normalized vacuum expectation value (vev)
\be
\label{knotvev}
W_R(\CK)={1\over Z(M)} \int {\mathcal D} A \, \re^{\ri S(A)} \, \tr_R \, U_\CK
\ee
is an invariant of the framed knot $\CK$, and it agrees with the quantum group invariant of $\CK$ based on the group $G$ and representation $R$. This 
can be extended to framed links $\CL$ of $L$ components $\CK_1, \cdots, \CK_L$ and in representations $R_1, \cdots, R_L$, 
\be
\label{linkvev}
W_{R_1, \cdots, R_L} (\CL)={1\over Z(M)} \int {\mathcal D} A \, \re^{\ri S(A)} \, \tr_{R_1} \, U_{\CK_1} \cdots \tr_{R_L} \, U_{\CK_L}
\ee
The invariant (\ref{linkvev}) turns out to be the quantum group invariant of the link $\CL$, based on the group $G$, and with ``colorings" $R_1, \cdots, R_L$. Our normalization is the natural one from the QFT point of view, namely, the invariant of the unknot with coloring $R$ is the quantum dimension 
of $R$:
\be
W_R(\unknot)={\rm dim}_q R.
\ee
 In particular, for $G=U(N)$, the QFT vev (\ref{linkvev}) is the colored HOMFLY invariant of $\CL$, and will be denoted by $\CH_{R_1, \cdots, R_L} (\CL)$. It is a 
 rational function of the variables
\be
\label{qvars}
q=\re^{g_s/2}, \qquad \nu=q^N. 
\ee
 When $G=SO(N)$, (\ref{linkvev}) is the colored Kauffman invariant of $\CL$, and will be denoted by $\CG_{R_1, \cdots, R_L} (\CL)$. It is also a rational function 
 of two variables $q, \nu$, where now $q=\nu^{N-1}$. Our conventions are such that
 \be
 \CG_{\tableau{1}}(\ununknot)=1+{\nu -\nu^{-1} \over q-q^{-1}}. 
 \ee
Invariants based on $Sp(N)$ are equivalent to the invariants based on $SO(N)$, as we will review below. 

One of the most fascinating aspects of this QFT re-interpretation of knot and three-manifold invariants is that one can analyze the vevs (\ref{linkvev}) by using different 
techniques in QFT, and very often this leads to properties of the invariants which are not obvious at all from their 
mathematical definition. Let us give a few examples:

\begin{enumerate}
\item The semiclassical analysis of the path integral defining $Z(M)$ leads to an asymptotic expansion of the exact quantum invariant at small $g_s$. This 
relates combinatorial and geometric data of $M$ and leads to a conjectural asymptotics of the WRT invariant which has not been proved yet. 

\item The Chern--Simons QFT can be solved exactly by using its relationship with conformal feld theory. This is the point of view put forward by Witten in \cite{witten} and leads to the standard properties of quantum knot invariants, like for example skein relations. 

\item The perturbative analysis of (\ref{knotvev}) makes contact with the theory of Vassiliev invariants. In particular, different choices of gauge to compute (\ref{knotvev}) lead to different formulations of Vassiliev theory, see \cite{labastida} for a detailed review. 
\end{enumerate}

In this talk I will review the results which have been obtained in Chern--Simons theory by using another important QFT technique, namely the $1/N$ expansion \cite{thooft}. 
This approach can be used when the gauge group is a classical gauge group: $G=U(N)$, $SO(N)$ and $Sp(N)$. Most importantly, 
when described in this way, Chern--Simons theory is equivalent to a (topological) string theory, and this provides surprising relationships 
between quantum group invariants of knots and links, and Gromov--Witten invariants. These ideas first took form ten years ago, and I will attempt here to 
provide a (biased) overview of the subject. In section 2 I explain some QFT aspects of the $1/N$ expansion, and I 
motivate the string/gauge theory equivalence from a diagrammatic point of view. In section 3, after stating 
the precise conjectures for the large $N$ string dual description of Chern--Simons theory in the $U(N)$ case, 
I review some of its tests involving knots and links. In section 4 I focus on indirect tests based on integrality 
properties. Finally, in section 5, I explain in some detail the recent generalization of these integrality properties to the $SO/Sp$ case, i.e. to the Kauffman invariant 
of links. Section 6 contains some concluding remarks.

\section{The $1/N$ expansion}
\begin{figure}
\includegraphics[height=.8cm]{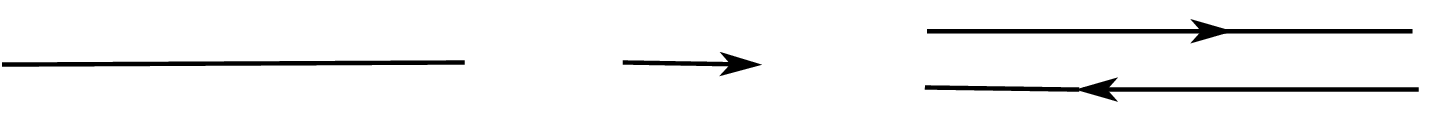}
\caption{Thickening an edge.}
\label{edgepropa}
\end{figure}
\begin{figure}
\includegraphics[height=5.7cm]{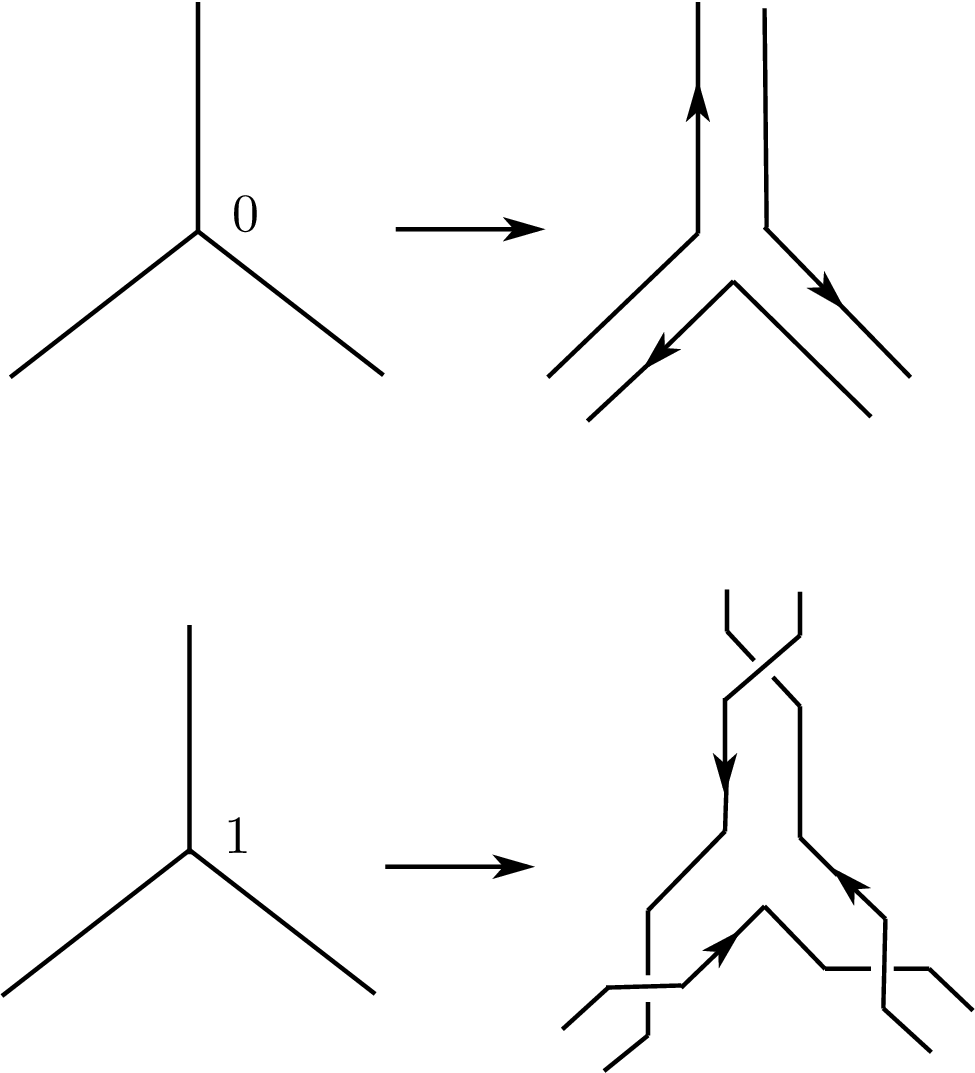}
\caption{Thickening a marked vertex.}
\label{resolution}
\end{figure}
The $1/N$ expansion was invented by 't Hooft in \cite{thooft}, and it can be used in any gauge theory with a classical 
gauge group. The general idea is to reorganize the diagrams of standard perturbation theory according to their power of $N$. 
The group factor of a Feynman diagram is in general a polynomial in $N$, so this suggests that 
we should ``split" each diagram in different components, each of them with a well-defined power of $N$. This leads to the double-line diagrams or 
fatgraphs of 't Hooft.

Let us now formalize this, in the case of $G=U(N)$, by using a well-known technique to Chern--Simons practicioners, namely the description of the 
$U(N)$ weight system given in \cite{cvitanovic,barnatan}. Let us consider the diagrams that appear in the calculation of the LMO invariant. These are 
trivalent diagrams modulo the IHX relations, and their space is usually denoted by $\CA(\emptyset)$. The $U(N)$ weight system is a map
\be
\label{wsystem}
W_{U(N)}:\CA(\emptyset) \rightarrow \IZ[N].
\ee
In order to construct $W_{U(N)}$, one first maps a trivalent diagram $D$ into a formal sum of {\it fatgraphs} or double-line diagrams, 
which are in fact Riemann surfaces of genus $g$ with $h$ boundaries. 
This formal sum is obtained as follows: first, we thicken the edges as shown in \figref{edgepropa}. Then, we mark all trivalent vertices of $D$ by $0$ or $1$. 
We thicken the marked vertices as shown in \figref{resolution}. We then obtain 
\be
\label{dmap}
D\rightarrow \sum_M (-1)^{s_M} \Sigma_{D,M}
\ee
where $M$ is the set of all possible markings, $s_M$ is the sum, over all vertices, of the values of $M$, and $\Sigma_{D,M}$ is the resulting fatgraph. 
Each fatgraph leads to a factor $N^{h(\Sigma_{D,M})}$, where $h(\Sigma_{D,M})$ is the number of boundaries of the Riemann surface. In terms of this construction, the weight system is given by
\be
W_{U(N)}(D)=\sum_M (-1)^{s_M} N^{h(\Sigma_{D,M})}.
\ee
This procedure just mimicks the structure of the Lie algebra of $U(N)$ \cite{cvitanovic,barnatan}: edges are associated to Lie algebra generators $T^a$. These are Hermitian 
matrices with two indices: one index in the fundamental representation, and another index in the anti-fundamental representation. These two indices lead to the two lines in \figref{edgepropa}, 
with opposite orientation. The vertices are associated to the structure constants of the Lie algebra, and the appearance of two markings in the vertices correspond to the two terms in the commutator 
of the Lie algebra. 
 
\begin{figure}
\includegraphics[height=2cm]{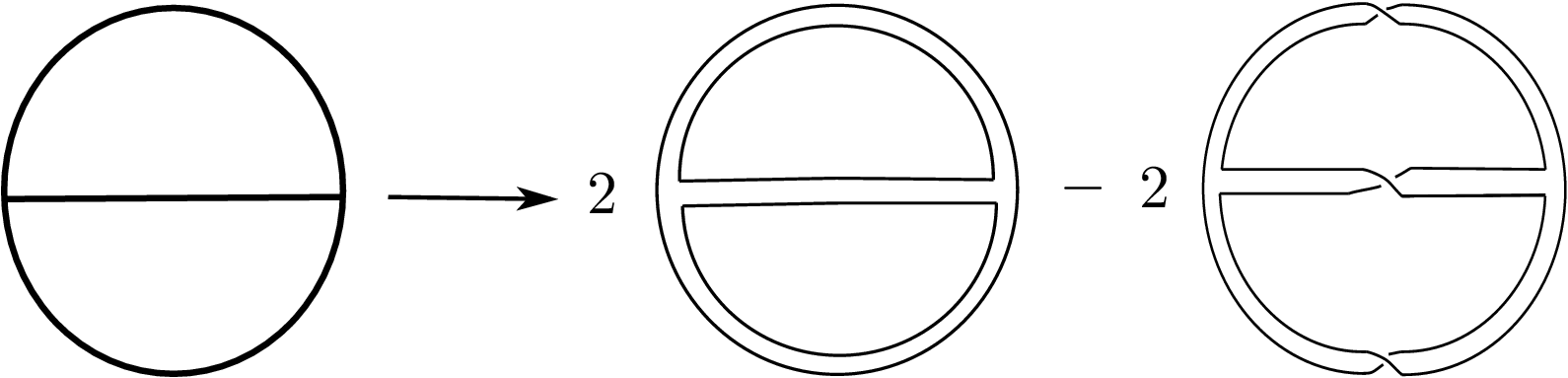}
\caption{The two Riemann surfaces with boundaries associated to a simple trivalent graph.}
\label{thetares}
\end{figure}

A simple example of the map (\ref{dmap}) is the theta graph, which leads to two different Riemann surfaces. One of them (the first one shown in \figref{thetares}) has $g=0$, $h=3$, while the second one has $g=1$, $h=1$. The trivalent fatgraphs of genus $g$ and with $h$ boundaries can also be regarded, by looking at 
the dual diagram, as triangulations of a {\it closed} Riemann surface of genus $g$. This is in fact the original picture in \cite{thooft}.

Let us now consider the expansion of Chern--Simons theory around the trivial flat connection, and let us denote by $Z$ the resulting partition function. By elementary principles of QFT, the {\it free energy} $F=\log Z$ has a perturbative expansion given as a sum over connected vacuum graphs, which are the trivalent 
graphs of $\CA(\emptyset)$. When this sum is reorganized in terms of fatgraphs, in the way we have explained, it has the structure
\be
\label{largenf}
F=\log \, Z=\sum_{g=0}^{\infty} F_g(t) g_s^{2g-2}
\ee
where
\be
t=N g_s
\ee
is the so called {\it 't Hooft parameter}. In this expression, $F_g(t)$ is in principle a formal power series, 
and it is obtained by summing the contributions of all fatgraphs of fixed genus $g$, weigthed by the power $t^h$, where $h$ is the number 
of boundaries of the fatgraph. For example, the first graph in \figref{thetares} gives a factor $g_s N^3=g_s^{-2} t^3$, with the power of $g_s$ 
appropriate for a fatgraph of genus zero. The expansion (\ref{largenf}) is called the {\it $1/N$ expansion of the free energy}, since $g_s=t/N$ and the resulting 
series at fixed $t$ can be also regarded as a series in $1/N$. The leading contribution to this expansion (i.e. for small $g_s$, or equivalently large $N$) 
comes from the fatgraphs of genus zero, also known as {\it planar graphs}. The quantities $F_g(t)$ are called {\it genus $g$ free energies} of the manifold $M$. 

The formal power series (\ref{largenf}) in two variables can be defined 
rigorously by using the LMO invariant as its starting point, as in \cite{glm}. Notice that 
we have restricted ourselves to the contribution of the trivial connection. The $1/N$ expansion for generic flat connections and for the full 
WRT invariant is more subtle and we will not discuss it here, see \cite{akmv,mpp} for some results in that direction in the case of lens spaces. 

So far, the genus $g$ free energies $F_g(t)$ are just formal power series. In the case of $M=\IS^3$ they can be computed in closed form for all $g$ \cite{gv}, and the resulting power series can be resummed to an analytic function at $t=0$. For example, for $g=0$ one has 
\be
\label{genuszero}
F_0(t) ={\rm Li}_3(\re^{-t}) -{1\over 2}t^2 \log t,
\ee
up to a polynomial of degree $3$ in $t$.

In general, perturbative series in QFT have zero radius of convergence. There are two reasons for such a behaviour:  the factorial growth of the number of Feynman diagrams contributing at each loop (see \cite{bender} for a review), and the factorial behavior of some class of diagrams (renormalons). 
Using the theory of the LMO invariant, it is possible to show that 
indeed there are no renormalons in Chern--Simons theory, as expected in a superrenormalizable QFT  \cite{gl}. 
But the perturbative series around the trivial flat connection is still factorially divergent due to the growth in the number of diagrams. 
It is also expected that the number of fatgraphs of genus $g$ grows only exponentially \cite{nussinov}, so that in a theory without renormalons 
quantities like $F_g(t)$ are analytic at the origin, for all $g$. This expectation from QFT at large $N$ can be actually 
rigorously proved for Chern--Simons theory, by using the theory of the LMO invariant, and one has the following
\begin{theorem} \cite{glm} $F_g(t)$ are analytic at $t=0$ for $g\ge 0$.
\end{theorem}

The $1/N$ analysis can be also applied to vevs of Wilson loops. In general, in order to have a well-defined $1/N$ expansion, 
one has to consider {\it connected} vevs, and this involves considering complicated combinations of Wilson loops in different representations (see 
for example \cite{lm,lmv}). The simplest example is the Wilson loop for a knot in the fundamental representation. In this case, the connected vev is equal to the ordinary vev, and one finds the 
large $N$ structure
\be
\label{fundN}
\CH_{\tableau{1}}(\CK)=\sum_{g\ge 0} g_s^{2g-1} \CH_g^{\CK}(t). 
\ee
The quantities $\CH_g^{\CK}(t)$ also have a diagrammatic representation. The relevant space of diagrams is $\CA(\IS^1)$, i.e. trivalent diagrams with a single 
$\IS^1$ boundary. The map (\ref{wsystem}) leads to fatgraphs which can be interpreted as triangulations of Riemann surfaces with one boundary, 
corresponding to the $\IS^1$. The formal power series in $t$ $\CH_g^{\CK}(t)$ is then the contribution of fatgraphs with fixed genus $g$ and one boundary.

\begin{figure}
\includegraphics[height=1.4cm]{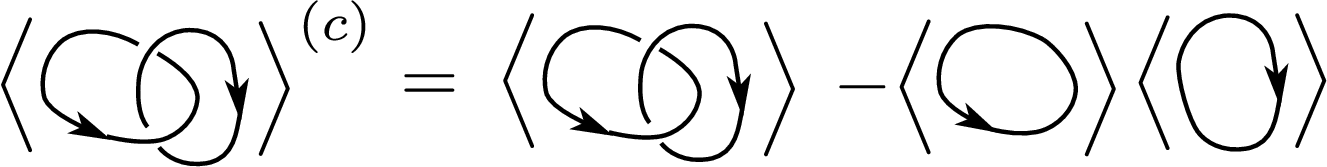}
\caption{Connected vev for a two-component link.}
\label{conlink}
\end{figure}

The case of links where all components are in the fundamental representation is also relatively simple and important. For a two-component link $L=2$, 
one has
\be
\CH^{(c)}_{\tableau{1}, \tableau{1}}(\CL)=\CH_{\tableau{1}, \tableau{1}}(\CL)-\CH_{\tableau{1}}(\CK_1)\CH_{\tableau{1}}(\CK_2).
\ee
This is illustrated in \figref{conlink} in the case of the Hopf link. In general, the connected vev of a link in the 
fundamental representation is obtained by subtracting to the original vev all possible products of vevs of sublinks, 
with signs \cite{lmv}. The combinatorics is the same one that appears in the cumulant expansion of probabilities. The $1/N$ expansion of these vevs is of the 
form 
\be
\label{Lexpansion}
\CH^{(c)}_{\tableau{1}, \cdots, \tableau{1}}(\CL)=\sum_{g\ge 0} g_s^{2g-2+L} \CH_g^{\CL}(t). 
\ee
This structure is a simple consequence of the diagrammatic expansion, and it has appeared in the work of Przytycki and others on the Vassiliev 
theory of Brunnian links (see for example \cite{pr}).  

Of course, one can construct more general connected vevs for knots and links, and they involve nontrivial combinations of the standard vevs 
in representations $R$ 
(which are in turn given by quantum group invariants). Therefore, the first lesson of the $1/N$ expansion is that the natural invariants of knots and links are not 
the vevs (\ref{knotvev}), (\ref{linkvev}), but rather their connected versions. 

\section{The $1/N$ expansion and string theory}

\subsection{The gauge theory/string theory correspondence}

The appearance of Riemann surfaces in the $1/N$ expansion has been regarded as an indication that gauge theories, when reorganized in this way, are described by string theories. We will refer to this idea as the {\it gauge theory/string theory correspondence or duality}. The correspondence was already suggested in the seminal paper by 't Hooft \cite{thooft}, but for a long time it remained a rather speculative idea due to the lack of concrete examples. The situation changed in the early nineties, where some simple string theories were found to be described by matrix models in the $1/N$ expansion \cite{dfgz}. But the true turning point was the AdS/CFT correspondence of Maldacena, which postulated a duality between $\CN=4$ super Yang--Mills theory and type IIB string theory on AdS$_5\times \IS^5$ \cite{malda}. In the AdS/CFT correspondence, the 't Hooft parameter corresponds to the common radius of the AdS$_5$ and the five-sphere, i.e. to a geometric parameter of the target space for the string. 

A very powerful aspect of the gauge/string theory correspondence is that there is a precise dictionary between gauge theory quantities and string theory quantities, and very often a difficult calculation in gauge theory gets translated into a simple calculation in string theory, and viceversa. For example, 
the quantities $F_g(t)$, i.e. the free energies of the gauge theories in the $1/N$ expansion on a compact manifold, should have an interpretation as 
free energies at genus $g$ of a string theory. Another general expectation is that $\CH_g^{\mathcal K}(t)$, i.e. the genus expansion of a Wilson loop vev in the fundamental representation around a contour $\CK$, should correspond to an amplitude in {\it open} string theory at genus $g$ and with one boundary. 
The boundary conditions of the open string should be fixed by the geometry of the loop $\CK$. Wilson loop vevs involving higher representations 
should correspond to Riemann surfaces with many boundaries. These ideas have been tested in the AdS/CFT correspondence in great detail. In view of 
the success of this idea, a natural question is: what is the string theory description of the $1/N$ expansion of Chern--Simons theory?

\subsection{String theory and Chern--Simons theory}

Building on \cite{wittencs}, Gopakumar and Vafa proposed in 1998 a conjectual string theory description of Chern--Simons theory on $\IS^3$ \cite{gv}. The 
relevant string theory is type A topological string theory (see for example \cite{ak} for more details), and its target is the simplest (non-compact) Calabi--Yau manifold, namely the {\it resolved conifold}
\be
\label{rescon}
X=\CO(-1)\oplus \CO(-1) \rightarrow \IP^1. 
\ee
The 't Hooft parameter $t$ is identified with the complexified K\"ahler parameter of $X$, i.e. with the complexified area of the $\IP^1$ base. 
The conjecture of Gopakumar and 
Vafa can be checked by simply computing the free energies of the topological string theory. These are known to be generating functionals of Gromov--Witten invariants $n_{g,d}$ of $X$, 
\be
F_g(t)=\sum_{d\ge 1} n_{g,d} \re^{-dt}. 
\ee
The invariant $n_{g,d}$ ``counts" in an appropriate sense the number of holomorphic curves of genus $g$ and degree $d$ in $X$. In the case of genus zero, 
these curves are just multicoverings of $\IP^1$, and the Aspinwall--Morrison formula \cite{am} gives
\be
n_{0,d}={1\over d^3}. 
\ee
We then find,  
\be
F_0(t)=\sum_{d\ge 1} {\re^{-dt}\over d^3}={\rm Li}_3 (\re^{-t}). 
\ee
This agrees with the calculation in (\ref{genuszero}). More precisely, the full agreement involves adding a term $t^2 \log\, t/2$ to (\ref{genuszero}), which indeed comes from the overall measure of the Chern--Simons path integral which we have not taken into account in the calculation. An explicit determination 
of the $F_g$ for $g>0$ confirms 
indeed that the 't Hooft resummation of genus $g$ fatgraphs in Chern--Simons theory on $\IS^3$ agrees with the generating function of genus $g$ Gromov--Witten invariants (see for example \cite{ak} for a detailed exposition). 

This is a remarkable agreement, but it only involves a simple quantity, namely the total free energy. 
This gauge/string correspondence can be considerably enriched by incorporating Wilson loops, as first done by Ooguri and Vafa in \cite{ov}. 
Their conjectural correspondence can be stated as follows. Given a link $\CL$ in $\IS^3$, there exists a Lagrangian submanifold $L_\CL$ 
of $X$ providing boundary conditions for open topological strings. The genus expansion of the connected vevs of Wilson loops based on this link are generating functionals for open Gromov--Witten invariants, counting holomorphic maps from Riemann surfaces with boundaries to $X$.  The boundaries map to $L_\CL$. 

\begin{figure}
\includegraphics[height=4cm]{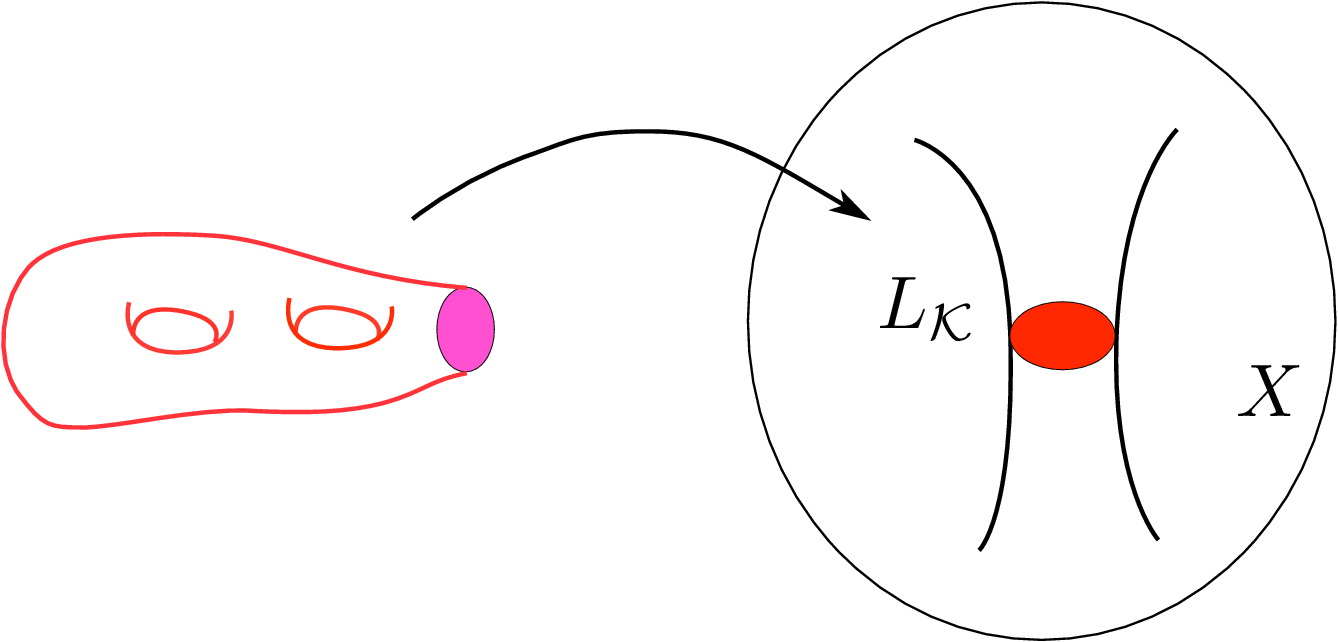}
\caption{In the string theory dual, the HOMFLY polynomial of a knot $\CK$ is interpreted in terms of maps of a Riemann surface with one boundary to the resolved conifold $X$. The boundary 
is mapped to a Lagrangian submanifold $L_\CK$.}
\label{mapknot}
\end{figure}
At this stage, this conjecture is not very precise, but it provides a far-reaching connection between knot invariants and Gromov--Witten invariants. Consider for example a knot $\CK$ in $\IS^3$, and the Wilson loop vev $\CH_{\tableau{1}}(\CK)$. As we explained in (\ref{fundN}), 
this vev (which is nothing but the HOMFLY polynomial of $\CK$, in the normalization of \cite{witten}) 
has a $1/N$ expansion involving functions $\CH_g^{\CK}(t)$, where $t$ is the 't Hooft parameter. 
According to \cite{ov}, this function is a generating functional of the 
form 
\be
\CH_g^{\CK}(t) =\sum_Q n_{g,Q}^{\CK} \, \re^{Q t}
\ee
where $n_{g,Q}^{\CK}$ are now {\it open} Gromov--Witten invariants. They count holomorphic maps from Riemann surfaces with 
one boundary which maps to the Lagrangian 
submanifold $L_\CK$, see \figref{mapknot}.

Clearly, in order to be more precise we need various ingredients, like for example a detailed construction of the map
\be
\label{LGmap}
\CL \subset \IS^3 \rightarrow L_\CL \subset X.
\ee
We also need a precise definition of open Gromov--Witten invariants. We will now discuss these issues, in the more general context of tests of the correspondence.

\subsection{Tests of the correspondence}

Since the Gopakumar--Vafa correspondence was proposed, there have been some non-trivial tests involving knots and links. These are the following:

\begin{enumerate}
\item For the (framed) {\it unknot} $\CK=\unknot$, there is a candidate Lagrangian submanifold \cite{ov}, and open Gromov--Witten invariants can be defined and calculated in terms of integrals on the Deligne--Mumford moduli space $\overline M_{g,n}$ \cite{kl,ls}. The correspondence leads to a highly nontrivial formula expressing these integrals in terms of quantum 
dimensions \cite{mv}, which generalizes the celebrated ELSV formula for Hurwitz numbers. This formula is by now well-established \cite{llz,op}. 
It is interesting to point out that 
the simplest case of this correspondence, at the level of knots, produces already an amazing amount of information!  The Hopf link can be also understood in this framework \cite{llztwo}. 

\item For generic knots and links there are two different proposals for the map (\ref{LGmap}). The construction of \cite{lmv,taubes}, based on twistor ideas, 
only applies to a special class of knots. 
There is another, more general proposal, which is quite natural \cite{koshkin}. It is not clear how these two proposals are related. Unfortunately, the corresponding open Gromov--Witten 
invariants have not been defined or computed, even at the heuristic level, and very little progress has been made in the last ten years based on direct computations of knot invariant in the Gromov--Witten theory side. 

\end{enumerate}

One can also try to consider extensions of the Gopakumar--Vafa conjecture which are amenable to test. There have been three main generalizations:

\begin{enumerate} 

\item So far the discussion has involved only Chern--Simons theory with a $U(N)$ gauge group. One can consider other classical gauge groups, like $SO(N)$ (after all, this leads to well-studied link invariants, 
like the Kauffman polynomial \cite{k}) and study their large $N$ dual. We will come back to this later on. 

\item Another possibility is to consider Chern--Simons theory on more general three-manifolds. The lens spaces $L(p,1)$ are relatively well-understood, and the string/gauge theory correspondence can be extended to these backgrounds \cite{akmv}. 

\item One can consider refinements of the standard Chern--Simons invariants, like those coming from Khovanov homology \cite{khovanov}. It has been conjectured that this refinement 
has a string theory counterpart \cite{gsv}, and some non-trivial tests of this conjecture have been made for the Hopf link in \cite{gikv}. 

\end{enumerate}

In this talk I will concentrate on an {\it indirect} test of the correspondence, namely the integrality properties of link invariants derived from the string theory picture. These properties 
were conjectured in \cite{ov,lmv} for the invariants associated to the $U(N)$ gauge group, and I will review them here. I will also review these properties in the $SO(N)/Sp(N)$ case, 
which were partially conjectured in \cite{bfmtwo} and finally spelled out in detail in \cite{mk}.

\sectiono{Integrality properties}

The basic idea behind the integrality tests of the Gopakumar--Vafa correspondence is the following. The correspondence states 
that link invariants can be expressed in terms of 
open Gromov--Witten invariants. If these open invariants enjoy {\it structural properties} which are not manifest in the link invariants, and if these properties turn out to be true in the 
knot theory side, we have provided an indirect test of the conjecture. 

What these properties could be? There is strong evidence that generating functionals of closed Gromov--Witten invariants, which are rational numbers, can be reexpressed in terms of 
{\it integer} invariants. These are called {\it Gopakumar--Vafa invariants} \cite{gvinv}, and there has been recent progress in defining them rigorously \cite{pt}. These integer invariants 
are interpreted as counting BPS states in compactifications of type II superstring theory. The physics arguments that led to the construction of the Gopakumar--Vafa invariants can be generalized to the open case \cite{ov,lmv}, and this leads indeed to conjectural, structural properties of knot and link invariants \cite{ov,lmv,lmtwo}. 

Let us first present a concrete example. Consider the vev $\CH_{\tableau{1}}(\CK)$ for an arbitrary knot $\CK$ in the fundamental 
representation of $U(N)$. The standard $1/N$ expansion leads to 
the representation (\ref{fundN}), where $\CH_g^{\CK}(t)$ are generating functions of open Gromov--Witten invariants. The integrality properties obtained with physics arguments in \cite{gvinv,ov,lmv} 
say that this vev can be written as 
\be
\label{simplepred}
\CH_{\tableau{1}}(\CK)=\sum_{g\ge 0}\sum_{Q\in \IZ} N_{\tableau{1};g,Q} (q-q^{-1})^{2g-1} \nu^Q, 
\ee
where the variables $q, \nu$ were defined in (\ref{qvars}), 
and $N_{\tableau{1};g,Q} $ are BPS {\it integer} invariants, labeled by two ``quantum numbers" $g,Q$ which are related to the genus and degree of 
embedded Riemann surfaces in the resolved conifold $X$. In this case, the predicted structure of $\CH_{\tableau{1}}(\CK)$ is well-known, and it follows from the skein relations for the 
HOMFLY polynomial. The BPS 
structural result ``explains" in some sense the appearance of this structure, and moreover it gives a heuristic interpretation of the integers appearing in the HOMFLY polynomials 
in terms of Euler characteristics of appropriate moduli spaces \cite{lmv}. This leads immediately to the possibility of refining these invariants by considering the cohomology of these 
spaces. In this sense, the BPS description of the HOMFLY polynomial should lead to a natural categorification. According to \cite{gsv}, this is nothing but the one proposed by 
Khovanov. 

The prediction (\ref{simplepred}) is only the simplest one in a relatively involved integrality conjecture formulated in \cite{ov,lmv,lmtwo}. Let us state this conjecture in the 
case of knots. Let $\CK$ be a knot, and let $\CH_R(\CK)$ be the $U(N)$ Chern--Simons invariant (\ref{knotvev}) in the representation $R$. 
We first define the generating functional 
\be
\label{zh}
Z_{\CH} (v) =\sum_R \CH_R(\CK) s_R(v),
\ee
understood as a formal power series in Schur polynomials $s_R(v)$. Notice that we are working formally in the limit $N=\infty$, so that all representations $R$ are allowed, and similarly, in constructing the Schur polynomials, we are considering the ring of symmetric polynomials in an infinite number of 
variables. In (\ref{zh}) we sum over all possible colorings, including the empty one $R=\cdot$. We now define the 
{\it reformulated HOMFLY invariants} of $\CK$, $f_R(q,\nu)$ through the equation
\be
\label{reformulated}
\log \, Z_{\CH}(v)= \sum_{k=1}^{\infty} \sum_R {1\over k} f_R(q^k, \nu^k) s_R(v^k),
\ee
where the l.h.s. is understood as a formal power series. One can easily prove \cite{lmtwo} that this equation determines uniquely the reformulated HOMFLY invariants $f_R$ in 
terms of the colored HOMFLY invariants of 
$\CK$. Explicit formulae for $f_R$ in terms of $\CH_R$ for representations with up to three boxes are listed in \cite{lmtwo}. 

Let $C_{\mu}$ be the conjugacy class of the symmetric group $S_{\ell(\mu)}$ associated to a partition $\mu$, 
where $\ell(\mu)$ is the lenght of the partition. We define
\be
z_\mu ={\ell(\mu)! \over |C_{\mu}|}.
\ee
If $\ell (R) =\ell (S)$, we define the matrix 
\be
M_{R S}=\sum_{\mu}  {1\over z_{\mu}} \chi_R(C_\mu) \chi_{S}(C_\mu) {\prod_{i=1}^{\ell(\mu)} \bigl( q^{\mu_i} -q^{-\mu_i} \bigr) \over q -q^{-1}} 
\ee
which is zero otherwise. It is easy to show that this matrix is invertible. We finally define
\be
\label{hattedf}
\hat f_R(q,\nu) =\sum_{S} M_{RS}^{-1} f_S(q, \nu). 
\ee
In principle, $\hat f_R(q,\nu)$ are rational functions, and they belong to the ring $\IQ[q^{\pm 1}, \nu^{\pm 1}]$ with denominators given by products of $q^r-q^{-r}$. However, we have the following 
\begin{conjecture} \label{conj} $\hat f_R(q,\nu) \in z^{-1} \IZ[z^2, \nu^{\pm1}]$, where 
\be
z=q-q^{-1} 
\ee
i.e. they have the structure
\be
\hat f_R(q,\nu) =z^{-1} \sum_{g\ge 0} \sum_{Q \in \IZ} N_{R; g,Q} z^{2g} \nu^Q,
\ee
where $N_{R;g,Q}$ are integer numbers and are called the BPS invariants of the knot $\CK$. The sum appearing here is finite, i.e. for a given knot 
and a given coloring $R$, the $N_{R;g,Q}$ vanish except for finitely many values of $g$, $Q$. 
\end{conjecture}

This conjecture extracts the BPS invariants from the standard quantum group invariants, and it has been recently proved in \cite{lp}. It can be extended to 
links, although the required notation is slightly more cumbersome. One simple consequence of the extension to links is that
\be
\label{lmovL}
\CH^{(c)}_{\tableau{1}, \cdots, \tableau{1}}(\CL)\in z^{L-2}\IZ[z^2, \nu^{\pm1}]
\ee
Notice that this implies (\ref{Lexpansion}), but it is much stronger. The conjecture (\ref{lmovL}) is maybe the simplest consequence of the string theory description of quantum 
group invariants. It agrees with, and generalizes, previous results by Lickorish--Millett \cite{lim} and Kanenobu--Miyazawa \cite{km} on the structure of the HOMFLY polynomial of links. In particular, it implies the Kanenobu--Miyazawa conjecture on the HOMFLY polynomial of Brunnian links.

\sectiono{Generalization to $SO/Sp$ gauge groups}

\begin{figure}
\includegraphics[height=1cm]{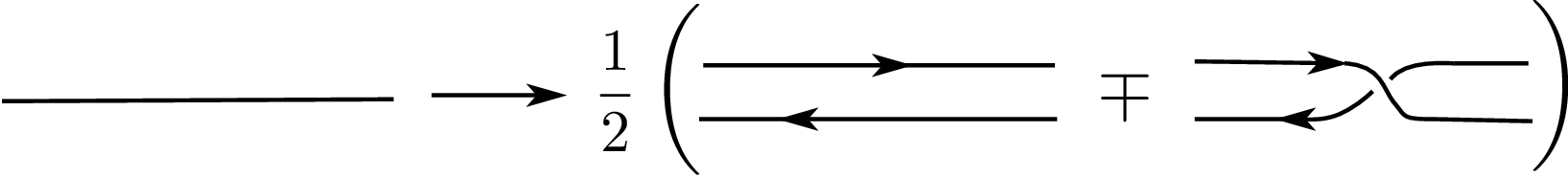}
\caption{Thickening an edge in the weight system of $SO/Sp$. The $\mp$ sign corresponds to the groups $SO$, $Sp$, respectively.}
\label{sosppropa}
\end{figure}

\begin{figure}
\includegraphics[height=5cm]{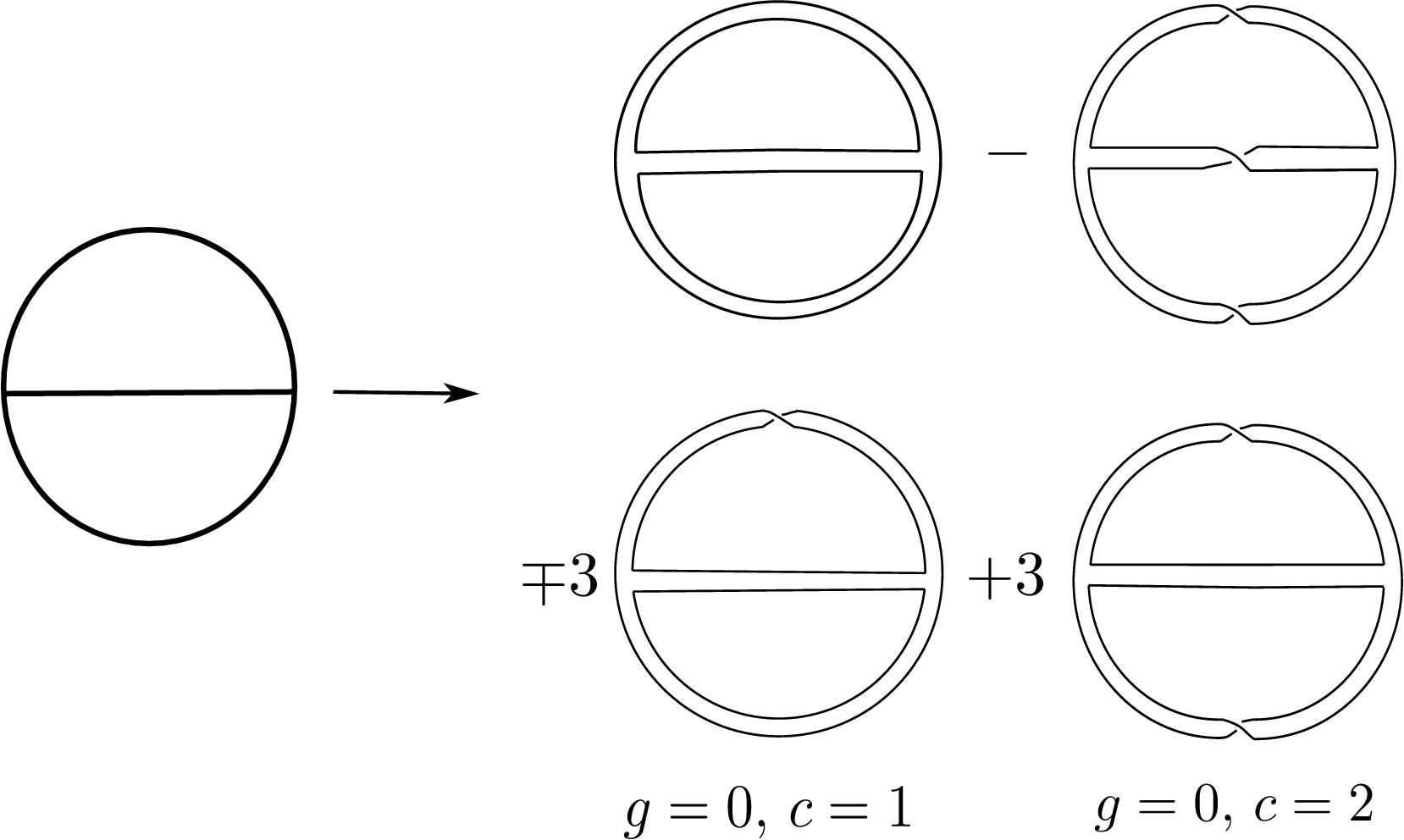}
\caption{The double-line diagrams obtained with the $SO/Sp$ weight system.}
\label{unortheta}
\end{figure}

The integrality conjecture of \cite{ov,lmv} (now a theorem) provides a strong indirect test of the string theory description of link invariants. It is an expected property of topological string amplitudes, which on the knot theory side is far from obvious. Moreover, it reproduces in an elegant way simple structural properties of 
the HOMFLY polynomial. In order to provide further tests of the string theory description, it would be very interesting to extend it to the other classical gauge groups. This is not merely an academic exercise, since the quantum group invariant based on $SO/Sp$, namely the colored Kauffman invariant of links, 
is definitely an important object in quantum topology. As we will see, there is such a string theory description, which in particular makes possible to ``explain" 
many mysterious properties of the Kauffman invariant, and in particular its relationships to the HOMFLY invariant. 

The best starting point to understand what is the appropriate string theory is to look again at the type of diagrams that appear in the weight system. In the 
$U(N)$ case, they indicated us the presence of oriented Riemann surfaces. However, in the $SO/Sp$ case, the weight system involves in a crucial way 
{\it non-orientable} Riemann surfaces \cite{cicuta}. This is due to the fact that the edges are thickened as shown in \figref{sosppropa}. 
Non-orientable surfaces are characterized topologically by their genus $g$ and the number of crosscaps $c$, which can be one or two. 
For example, the theta graph considered in \figref{thetares}, when evaluated with the weight system of $SO/Sp$, leads to the two orientable Riemann surfaces 
we had obtained before, together with two non-orientable 
Riemann surfaces with $c=1$ and $c=2$, respectively, as we can see in \figref{unortheta}. The $\mp$ sign 
in the $c=1$ diagram corresponds to the two different choices of gauge group $SO/Sp$. It follows from the rule to thicken the edges that 
the weight systems of $SO$ and $Sp$ differ just in the 
sign of the $c=1$ double-line diagrams. This leads in particular to the $SO(-N)=Sp(N)$ relationship between the $SO$ and the $Sp$ weight systems \cite{cvitanovic,takata}, and 
implies that the Chern--Simons invariants based on $SO(N)$ and $Sp(N)$ are equivalent up to a trivial change of sign in the $c=1$ contributions. For concreteness we will focus on 
the $SO(N)$ gauge group. 

Notice that the oriented diagrams appear in \figref{unortheta} with half the coefficient they had in \figref{thetares}. This is also a general feature of the 
$SO/Sp$ system, and we can write, schematically,
\be
\label{SOUcomp}
SO(N)/Sp(N)={1\over 2}U(N) +\text{non-orientable}.
\ee
One is then led to conjecture that, for a knot in the fundamental representation of $SO(N)$, we should have 
\be
\label{fundSO}
\CG_{\tableau{1}}(\CK)=\sum_{g\ge 0} g_s^{2g-1} \CG_{g,c=0}^{\CK}(t) +  \sum_{g\ge 0} g_s^{2g} \CG_{g,c=1}^{\CK}(t) +\sum_{g\ge 0} g_s^{2g+1} \CG_{g,c=2}^{\CK}(t), 
\ee
The first term gives the contribution of orientable surfaces, and it should be equal to the contribution appearing in (\ref{fundN}), i.e. 
\be
\CG_{g,c=0}^{\CK}(t) =\CH_{g}^{\CK}(t). 
\ee
The second and third terms in (\ref{fundSO}) are the contributions from non-orientable surfaces with $c=1$ and $c=2$ crosscaps, respectively. 
The counterpart of (\ref{SOUcomp}) is then
\be
\label{KHsimple}
\text{Kauffman}=\text{HOMFLY} + \text{non-orientable}
\ee

\begin{figure}
\includegraphics[height=4cm]{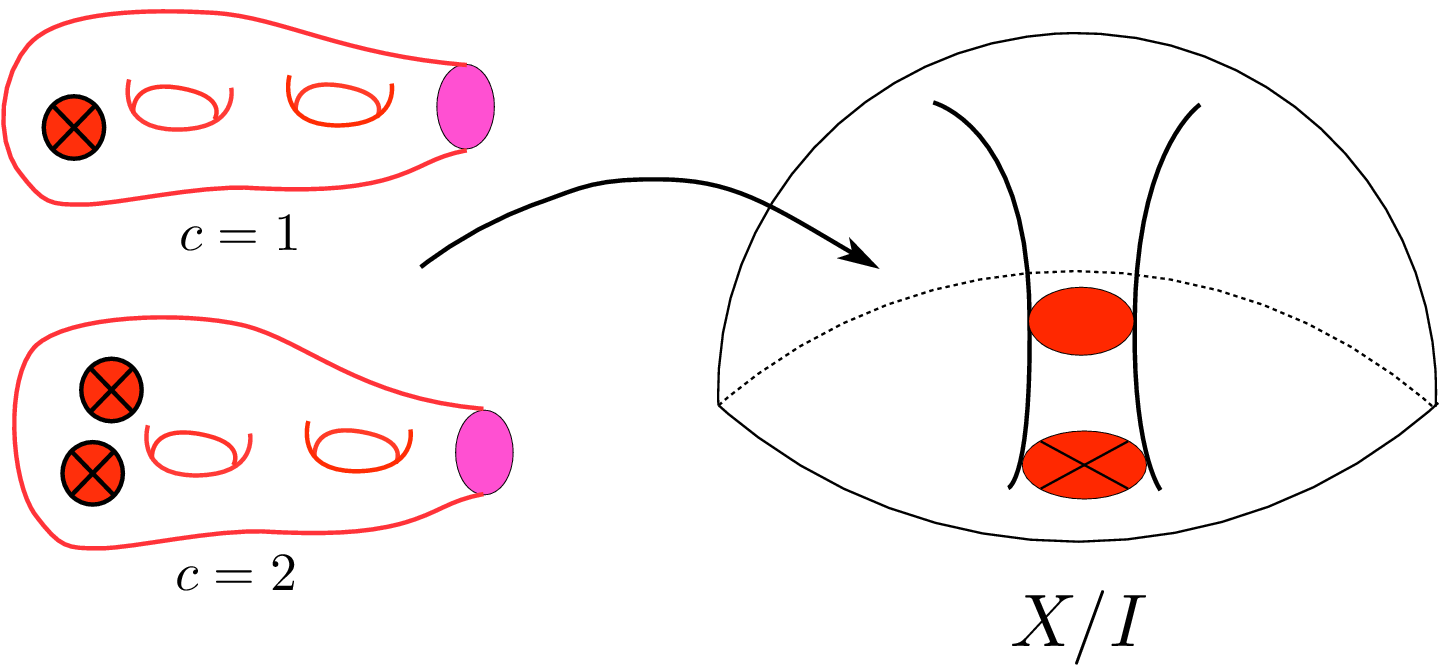}
\caption{In the string theory dual to $SO/Sp$ Chern--Simons theory, knot invariants have a non-orientable contribution given by maps of a Riemann surface with crosscaps to a 
quotient of the resolved conifold $X/I$.}
\label{mapinvo}
\end{figure}
According to the gauge/string theory correspondence, the contributions with $c\not=0$ should correspond to maps from non-orientable Riemann surfaces to a quotient of the 
resolved conifold $X$ by a free involution $I$. One convenient description of this manifold is as follows. The resolved conifold (\ref{rescon}) is a toric manifold, defined by the 
equation
\be
|X_1|^2 + |X_2|^2 -|X_3|^2 - |X_4|^2 =t
\ee
and a further quotient by a $U(1)$ action where the coordinates $(X_1, \cdots, X_4)$ have charges $(1,1,-1,-1)$. In this description, the free involution is 
\be
\label{orres}
\ba
I:X & \rightarrow X,\\
(X_1, X_2, X_3, X_4) &\rightarrow ({\overline X}_2, -{\overline X}_1,
{\overline X}_4, -{\overline X}_3).
\ea
\ee
Notice that the quotient geometry contains a projective plane $\IR\IP^2=\IP^1/I$. One can now consider holomorphic maps from Riemann surfaces with crosscaps to $X/I$ 
(as shown in \figref{mapinvo}), and count 
them in an appropriate way. The resulting theory should be a non-orientable version of Gromov--Witten theory. In string theory, such a construction is called an {\it orientifold}. 
It is a particular case of a $\IZ_2$ orbifold in which the discrete group acts as a orientation reversal on the worldsheet and as an involution $I$ on the target. 
The above orientifold construction was introduced in 
\cite{sv} in order to describe the large $N$ dual of Chern--Simons theory with $SO/Sp$ gauge groups. It was further studied in \cite{dfm,bfmone,bfmtwo}. 

Like before, direct tests of this string/gauge theory duality are difficult, 
but one could try to make an indirect test based on integrality properties. It is easy to generalize 
the construction of BPS invariants to the non-orientable case. The simplest situation occurs again for the $SO/Sp$ vevs in the 
fundamental representation, and we have
\be
\label{Kfstr}
\CG_{\tableau{1}}(\CK)=\sum_{Q}\sum_{g\ge 0} \left( N_{\tableau{1};g,Q} z^{2g-1}+ N^{c=1}_{\tableau{1};g,Q} z^{2g}  +N^{c=2}_{\tableau{1};g,Q} z^{2g+1}\right) 
 \nu^Q,
\ee
where $N^{c=1,2}_{\tableau{1};g,Q}$ are BPS invariants labelled, in addition to the quantum number $g$, by a quantum number $c$ which corresponds to the number of crosscaps. The BPS invariant $N_{\tableau{1};g,Q}$ 
agrees with the one appearing in the HOMFLY invariant (\ref{simplepred}). A simple consequence of the 
structural result (\ref{Kfstr}) is that the polynomial in $\nu^{\pm 1}$ which multiplies $z^{-1}$ is 
common to the HOMFLY invariant and the Kauffman invariant of a knot (in the fundamental representation). 
This is a well-known property of the Kauffman invariant \cite{lickorish}. 

We can now clarify two subtle issues concerning the comparison between the Kauffman and the HOMFLY invariants. First of all, in the comparison of the weight systems, we saw that the $U(N)$ 
system is multiplied by $1/2$. Where did this factor go in the comparison between the invariants, in (\ref{fundSO})? 
A second issue is the following: the Kauffman invariant is an invariant of {\it non-oriented} knots and links, while the HOMFLY invariant is an invariant of {\it oriented} knots and links. How are we going to compare both objects? It turns our that these two issues are related to each other. In fact, the right version 
of (\ref{KHsimple}) is rather
\be
\label{KHp}
\text{Kauffman}={1\over 2} \sum_{{\text {orientations}}} \text{HOMFLY} + \text{non-orientable}
\ee
i.e. there is a factor of $1/2$ inherited from (\ref{SOUcomp}), but in comparing both invariants we have to sum over all possible orientations of the link in the HOMFLY side. This produces in fact an invariant of non-oriented links, and the second issue is also resolved. In the case of a knot, there are only two possible orientations, and the HOMFLY invariants are 
equal, therefore the r.h.s. of (\ref{KHp}) gives back (\ref{KHsimple}). However, in the case of links the sum over possible orientations involves different invariants, as we will make more explicit in a moment. 

Some ingredients in the integrality conjecture for $SO/Sp$ link invariants were uncovered in \cite{bfmone,bfmtwo}. The final picture was found in \cite{mk}. One of the key ingredients of this conjecture, inspired by the results of Morton and Ryder \cite{mr}, is that the HOMFLY contribution to the 
Kauffman invariant involves in general {\it composite representations} of $U(N)$. In mathematical terms, this implies that we need the full HOMFLY skein of the annulus, which has been 
much developed in the work of Morton and collaborators (see for example \cite{morton}). Composite representations are labelled by two Young diagrams $(R,S)$, 
and they can be expressed in terms of tensor products $V\otimes {\overline W}$, 
where ${\overline W}$ are conjugate representations. The precise formula is 
\be
\label{tensorcomposite}
(R,S)=\sum_{ U, V,W} (-1)^{\ell(U)} N^R_{U V} N^S_{U^T W}\, (V\otimes {\overline W}),
\ee
and $N_R^{ST}$ Littlewood--Richardson coefficients. The HOMFLY invariant of a knot $\CL$ whose components are colored with the composite representations $(R_1,S_1)$, $\cdots$, 
$(R_L, S_L)$ will be denoted by 
\be
\CH_{(R_1,S_1),\cdots, (R_L, S_L)}(\CL). 
\ee
One has the property
\be
\label{reversal}
\CH_{(R_1, S_1), \cdots, (S_j, R_j), \cdots, (R_L, S_L)}(\CL) =\CH_{(R_1, S_1), \cdots, (R_j, S_j), \cdots, (R_L, S_L)}({\overline \CL}_j),
\ee
where ${\overline \CL}_j$ is the link obtained from the link $\CL$ by reversing the orientation of the $j$-th component. In particular, 
\be
\label{revH}
\CH_{(R, S)}(\CK)=\CH_{(S,R)}(\overline \CK), 
\ee
where $\CK$ is an oriented knot and ${\overline \CK}$ is the knot with opposite orientation. Therefore, summing over all possible composite representations of a link involves in a particular 
summing over all its possible orientations. 

We are now ready to state the integrality conjecture for the Kauffman invariants proposed in \cite{mk}. Like before, we will restrict ourselves to the case of knots, and we will make some indications 
in the case of links. We introduce the two generating functionals
\be
\ba
Z_{H \overline H}(v)&=\sum_{R, R_1, R_2} N_{R_1 R_2}^R \CH_{(R_1, R_2)}(\CK)s_R(v), \\
Z_{\CG} (v)&=\sum_R \CG_R(\CK) s_R(v).
\ea
\ee
Notice that, due to the property (\ref{revH}), both are invariants of non-oriented knots. We define $g_R$ through
\be
\label{grdef}
\log Z_{\CG}(v) -{1\over 2} \log Z_{H \overline H}(v)=\sum_{k \, {\rm odd}} \sum_R {1\over k} g_R(q^k, \nu^k) s_R(v^k).
\ee
Here the sum over $k$ is over all 
positive {\it odd} integers. The factor $1/2$ in the l.h.s. is closely related to the factor $1/2$ in (\ref{SOUcomp}). We also define, as in (\ref{hattedf}), 
\be
\hat g_R(q,\nu) =\sum_S M_{RS}^{-1} g_S.
\ee
The analogue of the conjecture \ref{conj} is 

\begin{conjecture} \label{noconj} $\hat g_R(q,\nu)  \in  \IZ[z, \nu^{\pm1}]$, i.e. they have the structure
\be
\hat g_R(q,\nu) = \sum_{g\ge 0} \sum_{Q \in \IZ} \left( N^{c=1}_{R; g,Q} z^{2g} +N^{c=2}_{R; g,Q} z^{2g+1}\right) \nu^Q,
\ee
where $N^{c=1,2}_{R; g,Q}$ are integers. They are the BPS invariants associated to the non-orientable case.  
\end{conjecture}

When $R=\tableau{1}$, we recover the result (\ref{Kfstr}). For representations with two boxes, we have that
\be
\ba
g_{\tableau{2}}&=\CG_{\tableau{2}}-{1\over 2}\CG_{\tableau{1}}^2 -{\CH}_{\tableau{2}} + {\CH}_{\tableau{1}}^2
-{1\over 2}\CH_{(\tableau{1}, \tableau{1})}, \\
g_{\tableau{1 1}}&=\CG_{\tableau{1 1}}- {1\over 2}\CG_{\tableau{1}}^2 -{\CH}_{\tableau{1 1}} +
{\CH}_{\tableau{1}}^2-{1\over 2}\CH_{(\tableau{1},\tableau{1})}.
\ea
\ee
The integrality conjecture \ref{noconj} implies in particular that
\be
\CG^2 _{\tableau{1}}(\CK)=\CH_{(\tableau{1},\tableau{1})}(\CK) , \qquad {\rm mod}\, \, 2.
\ee
This is an intriguing theorem due to Rudolph \cite{rudolph}, recently generalized by Morton and Ryder to higher representations  \cite{mr}, which 
finds a ``natural" explanation in the context of BPS integrality properties in string theory. The generalization of \cite{mr} also seems to follow from 
the conjecture. 

\begin{figure}
\includegraphics[height=1cm]{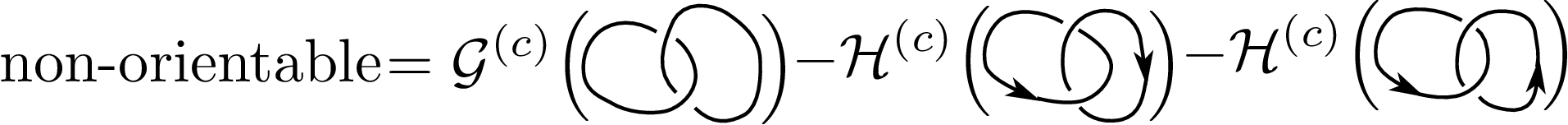}
\caption{In order to obtain the non-orientable contribution to the connected Kauffman invariant of a two-component, non-oriented link, one has to subtract the sum of the HOMFLY invariants for all possible orientations of the link.}
\label{nonorlink}
\end{figure}
The above integrality conjecture for $SO/Sp$ invariants can be extended to links. The simplest case of this conjecture says that
\be
\label{nonor}
\CG^{(c)}_{\tableau{1} , \cdots, \tableau{1}}(\CL) =\sum_{\alpha=1}^{2^{L-1}} \CH^{(c)}_{\tableau{1} , \cdots, \tableau{1}}(\CL_\alpha)+ g_{\tableau{1} , \cdots, \tableau{1}}(\CL), \qquad g_{\tableau{1} , \cdots, \tableau{1}}(\CL) \in z^{L-1}  \IZ[z, \nu^{\pm1}].
\ee
In this equation, the sum over $\alpha$ is over all possible non-equivalent orientations of the link $\CL$, denoted by $\CL_{\alpha}$ (two orientations are equivalent if they are related by 
an overall reversal of orientation of all the components in the link). If $\CL$ has $L$ components, there are $2^{L-1}$ non-equivalent orientations. The quantity $ g_{\tableau{1} , \cdots, \tableau{1}}(\CL)$ gives the non-orientable contribution to the invariant of the link, and as shown in (\ref{nonor}), it involves the sum over all possible orientations of $\CL$. The case of a two-component link is illustrated in \figref{nonorlink}. The conjecture 
(\ref{nonor}) implies some results of Kanenobu for the Kauffman invariant of links \cite{kanenobu}, and on top of that it leads to new relationships between the Kauffman polynomial of non-oriented links and the 
sum of the HOMFLY polynomials of all its possible orientations, see \cite{mk} for a detailed explanation and for examples. Further tests of the conjecture \ref{noconj} were performed in \cite{stevan,ramaK}. 

\sectiono{Conclusions}
The string theory description of Chern--Simons theory is extremely interesting, both from a physical and from 
a mathematical point of view. From a physical point of view, it provides one of the most 
detailed examples of a large $N$ string duality, a subject of paramount importance in modern theoretical physics. 
From a mathematical point of view, it provides a bridge between two major 
areas of modern mathematics. 

Unfortunately, progress in this area has been difficult and slow. Many important aspects of the 
correspondence still lack a sound mathematical basis, and there are no direct tests 
for nontrivial knots. Indirect tests based on integrality properties are so far our best indication that the conjectural string theory/Gromov--Witten 
description of Chern--Simons theory holds for 
general knots and links. One important ingredient in order to make further progress is to provide precise definitions of all the objects and invariants 
involved in the Gromov--Witten side.  

Our limited knowledge of this large $N$ duality has already given us very interesting results: new structural properties of quantum group invariants of knots and links, new results for intersection theory on the 
Deligne--Mumford moduli space, new perspectives on categorification... It is likely that further progress in this area will unveil even more treasures.

\bibliographystyle{amsalpha}

\end{document}